\documentclass[preprint2]{aastex}

\def\ltsima{$\; \buildrel < \over \sim \;$}
\def\gtsima{$\; \buildrel > \over \sim \;$}
\def\lsim{\lower.5ex\hbox{\ltsima}}
\def\gsim{\lower.5ex\hbox{\gtsima}}
\def\lapp{\ifmmode\stackrel{<}{_{\sim}}\else$\stackrel{<}{_{\sim}}$\fi}
\def\gapp{\ifmmode\stackrel{>}{_{\sim}}\else$\stackrel{<}{_{\sim}}$\fi}
\newcommand{\masyr}{${\rm mas\, yr^{-1}}$}
\newcommand{\masyrarc}{${\rm mas\, yr^{-1}\, arcsec^{-1}}$}

\usepackage{amsmath}
\usepackage{textcomp}
\usepackage{amssymb}

\newdimen\minuswidth    %define @ width of minus sign for tables
\setbox0=\hbox{$-$}
\minuswidth=\wd0
\catcode`@=\active
\def@{\kern\minuswidth}
\setbox0=\hbox{\rm0}
\shorttitle{Absolute proper motion in the direction of NGC~6681} 
\shortauthors{Massari et al.}
 
\begin{document} 

\title{\textit{HST} ABSOLUTE PROPER MOTIONS OF NGC 6681 (M70) AND THE
  SAGITTARIUS DWARF SPHEROIDAL GALAXY$^\ast$}

\author{
D.\ Massari\altaffilmark{1},
A.\ Bellini\altaffilmark{2},
F.\ R.\ Ferraro\altaffilmark{1},
R.\ P.\ van der Marel\altaffilmark{2},
J.\ Anderson\altaffilmark{2}, 
E.\ Dalessandro\altaffilmark{1},
and B.\ Lanzoni\altaffilmark{1}
}
\altaffiltext{1}{Dipartimento di Fisica e Astronomia, Universit\`a
  degli Studi di Bologna, v.le Berti Pichat 6/2, I$-$40127 Bologna,
  Italy}

\altaffiltext{2}{Space Telescope Science Institute, 3700 San Martin
  Drive, Baltimore, MD 21218, USA}
\altaffiltext{$^\ast$}{Based on archival observations with the
  NASA/ESA \textit{Hubble Space Telescope}, obtained at the Space
  Telescope Science Institute, which is operated by AURA, Inc., under
  NASA contract NAS 5-26555.}

%}

\date{\today}
%\email{davide.massari@unibo.it}

\begin{abstract} 
  We have measured absolute proper motions for the three populations
  intercepted in the direction of the Galactic globular cluster
  NGC~6681:\ the cluster itself, the Sagittarius dwarf spheroidal
  galaxy and the field. For this we used {\it Hubble Space Telescope}
  ACS/WFC and WFC3/UVIS optical imaging data separated by a temporal
  baseline of 5.464 years.  Five background galaxies were used to
  determine the zero point of the absolute-motion reference frame.
  The resulting absolute proper motion of NGC~6681 is
  ($\mu_{\alpha}\cos\delta, \mu_{\delta}$)=($1.58\pm0.18,
  -4.57\pm0.16$) \masyr.  This is the first estimate ever made for
  this cluster. For the Sgr dSph we obtain ($\mu_{\alpha}\cos\delta,
  \mu_{\delta})=(-2.54\pm0.18, -1.19\pm0.16$) \masyr, consistent with
  previous measurements and with the values predicted by theoretical
  models. The absolute proper motion of the Galaxy population in our
  field of view is ($\mu_{\alpha}\cos\delta,
  \mu_{\delta})=( -1.21\pm0.27, -4.39\pm0.26$) \masyr.  In this study
  we also use background Sagittarius Dwarf Spheroidal stars to determine 
  the rotation of the globular cluster in the plane of the sky and find
  that NGC~6681 is not rotating significantly:\ 
  $v_{\rm rot}=0.82\pm1.02$ km$\,$s$^{-1}$ at
  a distance of $1\arcmin$ from the cluster center.
\end{abstract}
 

\section{INTRODUCTION}

Galactic Globular Clusters (GCs) provide a powerful tool to
investigate the structure and the formation history of the Milky Way.
Indeed, they are fundamental probes of the Galactic gravitational
potential shape, from the outer region of the Galaxy (see
\citealt{casetti07}) to the inner Bulge (\citealt{casetti10}).  The
currently and most widely accepted picture for the formation of the
Galactic GC system (\citealt{zinn93}, \citealt{fb10}) points toward an
accreted origin for the outer ($r>10$ kpc) young halo (YH) GCs, while
a large number of the inner, old halo (OH) clusters probably formed
via dissipationless collapse, coevally with the collapse of the
protogalaxy.  The finding of several OH, metal-poor GCs with a thick
disc-like kinematics (\citealt{dinescu99}), sets a tight constraint on
the epoch of the formation of the Galactic disc.  Moreover, the
demonstration that several YH GCs are kinematically associated with
satellites of the Milky Way, such as the Sagittarius dwarf spheroidal
galaxy (hereafter Sgr dSph, see for instance \citealt{bellazzini03}),
gives important clues as to how the Galaxy was built up through merger
episodes.  The Sgr dSph (\citealt{ibata94}, \citealt{bfb99}) also provides one of the
best opportunities to study the shape, orientation and mass of the
Milky Way dark matter halo through investigation of its luminous
tidal streams. Recent studies have highlighted a so-called halo
conundrum (\citealt{law05}), showing that the available models were
not able to reproduce simultaneously the angular position, distance
and radial-velocity trends of leading tidal debris.  \cite{lm10a}
claim to have solved this conundrum by introducing a non-axisymmetric
component to the Galactic gravitational potential that can be
described as a triaxial halo perpendicular to the Milky Way disc. Even
if poorly motivated within the current Cold Dark Matter paradigm,
these findings have subsequently been confirmed by \cite{dw13}. 
However, \cite{debattista13} fail to reproduce
plausible models of disc galaxies using such a scenario.  In order to
make substantial progress towards a solution of this debate, new
observational data are needed, starting from accurate proper motions
(PMs).

The existence of other peculiar systems like $\omega$~Centauri
(\citealt{omega}) and Terzan~5 (\citealt{ferraro09}) harboring stellar
populations with significant iron-abundance differences
($\Delta$[Fe/H]$>0.5$ dex) supports a complex formation scenario for
the Galactic halo and the Bulge.  Therefore, a detailed description of
the kinematical properties of the Galactic GC system is a crucial
requirement to obtain new and stronger constraints on the formation
history of our Galaxy.  In this sense, publicly-available catalogs of
absolute PMs for several Galactic GCs are of great importance.  A
notable example is the ground-based Yale/San Juan Southern Proper Motion catalog
(\citealt{platais}, \citealt{dinescu97} and the following papers of
the series).
These kinds of studies are extremely difficult in regions of the sky where
different stellar populations overlap (such as towards the Bulge) and
the associated uncertainties are typically large, ranging between
$0.4$ \masyr and $0.9$ \masyr (\citealt{casetti07},
\citealt{casetti10}). In this sense the {\it Hubble Space Telescope}
({\it HST}) provides a unique opportunity to measure
high-accuracy stellar PMs even in the most crowded and complex regions
of the Galaxy, as seen in \cite{clarkson08}
or \cite{jay10}, for example.

\begin{figure}[!htb]
\includegraphics[width=\columnwidth]{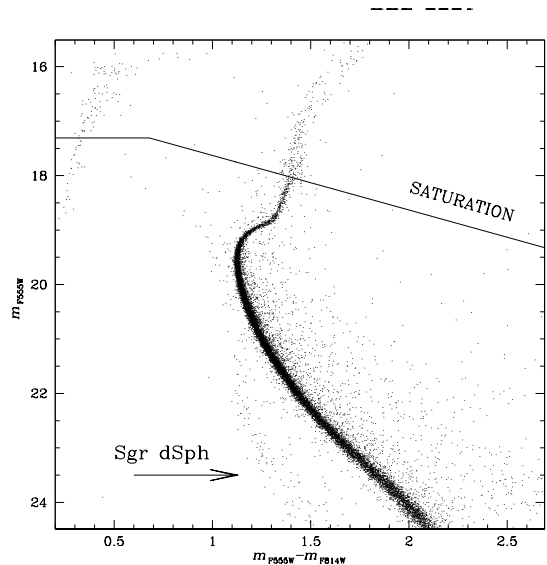}
\caption{\small The ({\it m}$_{{\rm F555W}}$, {\it m}$_{{\rm F555W}}-${\it
    m}$_{{\rm F814W}}$) UVIS/WFC3 CMD of NGC~6681. All of the
  evolutionary sequences of the cluster are well defined.  The
  broadening of the RGB and the HB is due to exceeding the saturation
  level, indicated in the plot.  The presence of the MS of the Sgr
  dSph at {\it m}$_{{\rm F555W}}>21$ mag and $1.1<${\it m}$_{{\rm
      F555W}}-${\it m}$_{{\rm F814W}}<1.8$ mag and the contribution of
  the field are also evident.}
\label{vvi}
\end{figure}

\begin{figure}[!htb]
\includegraphics[width=\columnwidth]{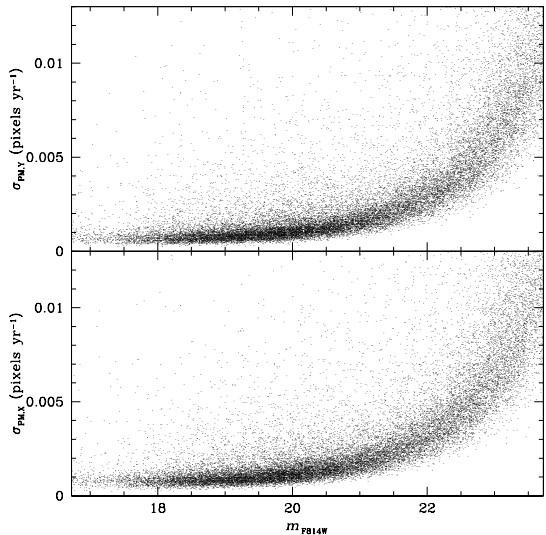}
\caption{\small Uncertainties, in units of pixel\, yr$^{-1}$, of the relative
  PMs along the Y-direction (upper panel) and the X-direction (lower
  panel), as a function of the {\it m}$_{{\rm F814W}}$ magnitude.  For
  stars brighter than {\it m}$_{{\rm F814W}}\sim20.7$ mag the bulk of
  the PM uncertainties are smaller than 0.002 pixel\, yr$^{-1}$.}
\label{err}
\end{figure}

As part of a project aimed at using blue straggler stars as tracers
of the dynamical evolution of GCs (\citealt[][2012]{ferrarom30}) we
obtained WFC3 observations of the globular cluster NGC~6681. In this
paper we present and discuss accurate PMs of stars in the cluster's
direction.  
This paper is part of, and uses techniques developed in the
context of, the HSTPROMO collaboration,\footnote{For details see
HSTPROMO home page at http://www.stsci.edu/~marel/hstpromo.html} a
set of HST projects aimed at improving our dynamical understanding
of stars, clusters, and galaxies in the nearby Universe through
measurement and interpretation of proper motions.
As it happens, NGC~6681 is located in an extremely interesting 
region of the sky:
it overlaps the main body of the Sgr dSph.  With the extraordinarily
high photometric and astrometric accuracy of {\it HST} we have
been able to separate the two populations and measure their
individual absolute PMs.  This is the first time that the PM of
NGC~6681 has been estimated.
 
The paper is structured as follows:\ in Section 2 we describe the {\it
  HST} data sets used for our investigation, providing a brief summary
of our data-reduction procedure;\ in Section 3 the relative PMs of our
sources are presented;\ in Section 4 we describe our method for obtaining
absolute PMs, by defining the zero-point reference frame and by
quantifying possible systematic errors;\ finally, in Section 5 we
present the results of our analysis.

\begin{figure*}[!htb]
\includegraphics[width=16cm]{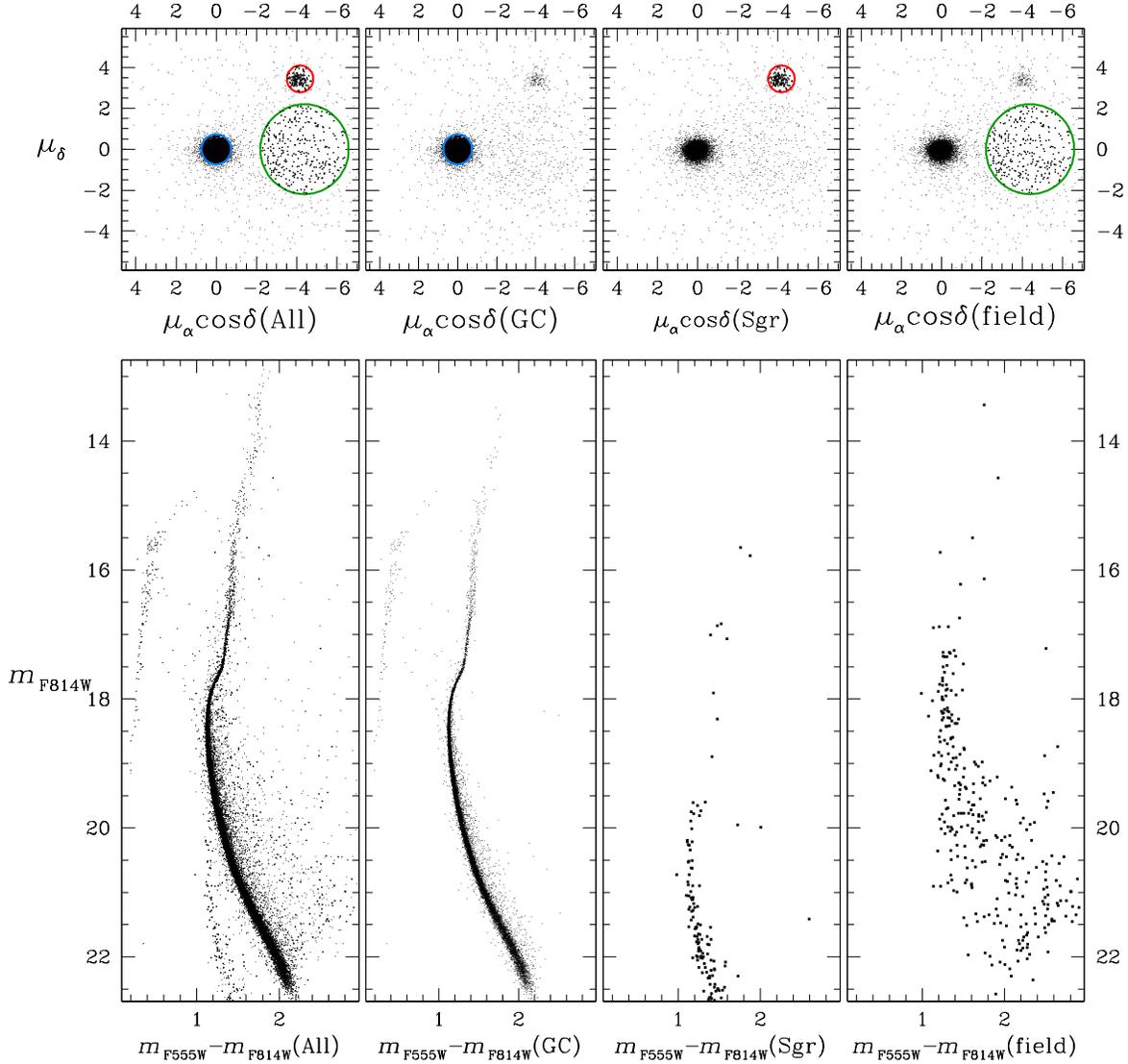}
\caption{\small The upper panels show the Vector Point Diagrams (VPDs)
  of the relative PMs. In the lower panels the CMDs corresponding to
  the selections applied in the VPDs are displayed. {\it First
    column}:\ in the VPD the different populations are indicated with
  different colors (a sample of cluster members in blue, of Sgr dSph
  stars in red, of the field in green), but no selection is
  applied. The corresponding CMD shows the entire PM catalog. {\it
    Second column}:\ in the VPD cluster members are selected within
  the blue circle and the corresponding CMD displays only well-defined
  cluster evolutionary sequences. {\it Third column}:\ Sgr dSph
  selection within the red circle and corresponding CMD.  {\it Fourth
    column}:\ the selection in the VPD (in green) of the bulk-motion
  of field stars and their location on the CMD.}
\label{relpm}
\end{figure*}

\section{OBSERVATIONS AND DATA REDUCTION}

In order to measure the PMs in the direction of NGC~6681 we used two
{\it HST} data sets.  The one used as first epoch was acquired under
GO-10775 (PI:\ Sarajedini). It consists of a set of high-resolution
images obtained with the Wide Field Channel (WFC) of the Advanced
Camera for Survey (ACS). The WFC/ACS is made up of two $2048 \times
4096$ pixel detectors with a pixel scale of $\sim0.05
\arcsec\,$pixel$^{-1}$ and separated by a gap of about 50 pixels, for
a total field of view (FoV) of $\sim200\arcsec \times 200\arcsec$.
For our investigation we used four deep exposures in both the F606W
and the F814W filters (with exposure times of $140$ sec and $150$ sec,
respectively), taken on May 20, 2006.  We work here exclusively with
the \_FLC images, which have been corrected with the pixel-based
correction in the pipeline (\citealt{jaybedin10}, and \citealt{ubedajay}).

The second-epoch data set is composed of proprietary data obtained
through GO-12516 (PI:\ Ferraro). This program consists of several
deep, high-resolution images taken with the UVIS channel of the Wide
Field Camera 3 (WFC3) in the F390W, F555W and F814W filters.  The WFC3
UVIS camera is made of two $2048 \times 4096$ pixel chips, separated
by a gap of approximately 30 pixels.  Its pixel scale is $\sim0.04
\arcsec\,$pixel$^{-1}$ and the total FoV is $162\arcsec \times
162\arcsec$.  The sample analyzed in this work consists of $9\times150\,$s
images in F555W and $13\times348\,$s images in
F814W. These images have not been corrected for CTE losses, since 
no pixel-based correction was available at the time of this reduction.  
These images were taken relatively soon after installation and background 
in these images is greater than 12 electrons, so any CTE losses should 
be small, particularly for the bright stars we are focusing on here 
(see \citealt{jayisr12}).
Since these observations were taken on November 5, 2011, the two data
sets provide a temporal baseline of $\sim 5.464$ yrs.

The data-reduction procedures are described in detail in
\cite{jayking06}. Here we provide only a brief description of the main
steps of the analysis.  The first step was to reduce each individual
exposure.  We analyzed the first epoch with the publicly available program
\texttt{img2xym$\_$WFC.09$\times$10}, which is documented in
\cite{jayking06}.  This program uses a pre-determined model of
spatially varying PSFs plus a single time-dependent perturbation PSF
(to account for focus changes or spacecraft breathing). The final output
is a catalog with instrumental positions and magnitudes for each
exposure.  We followed the same approach for the second epoch, by
using the program \texttt{img2xym$\_$wfc3uv}, which is 
similar to the ACS program described above.
We corrected the star positions in each catalog for
geometric distortion, by means of the solution provided by \cite{jayacs}
for the ACS camera and by \cite{bellini11} for the WFC3.

As a consistency check, we built the WFC3/UVIS ({\it m}$_{{\rm F555W}}$,
{\it m}$_{{\rm F555W}}-${\it m}$_{{\rm F814W}}$) CMD of NGC~6681.  The
F555W sample was constructed by selecting stars in common among at
least 4 out of 9 single-exposure catalogs, while the F814W 
sample is made
up of those stars detected in at least 5 out of 13 single-exposure
catalogs. The CMD resulting from these two samples is shown in Figure
\ref{vvi}. The instrumental magnitudes have been calibrated onto the
VEGAmag system using aperture corrections and zeropoints reported
in the WFC3 web
page\footnote{http://www.stsci.edu/hst/wfc3/phot\_zp\_lbn.}.  The CMD
exhibits well-defined cluster evolutionary sequences. The main
sequence (MS) extends to almost $5$ magnitudes below the Turn Off (TO)
region. Another bluer sequence is visible at {\it m}$_{{\rm
    F555W}}>21$ mag and $1.1<${\it m}$_{{\rm F555W}}-${\it m}$_{{\rm
    F814W}}<1.8$ mag and remains well separated from the cluster
MS. In the following sections we will demonstrate, by means of PM
membership, that it corresponds to the MS of the Sgr dSph (see
\citealt{siegel11}). The red giant branch (RGB) and the horizontal
branch (HB) of the cluster are almost entirely above the saturation
limit and thus are excluded from the PM analysis.  A more
detailed analysis of the stellar populations in NGC~6681 will be
presented in a forthcoming paper.

\begin{figure}[!htb]
\includegraphics[width=\columnwidth]{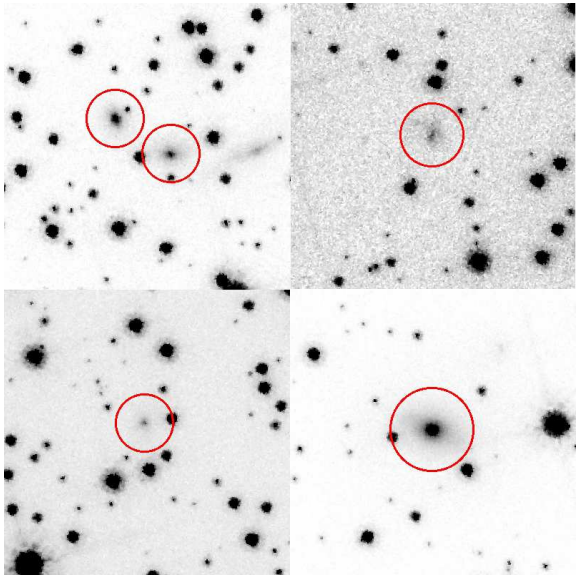}
\caption{\small The five selected background galaxies as they appear
  in the F814W images. They differ from the stellar sources since
  their light is more diffuse across the surrounding pixels. Their
  point-like nuclei allow us to accurately determine their centroid
  and thus to obtain a precise measure of their relative proper
  motions.}
\label{gal}
\end{figure}

\begin{figure}[!htb]
\includegraphics[width=\columnwidth]{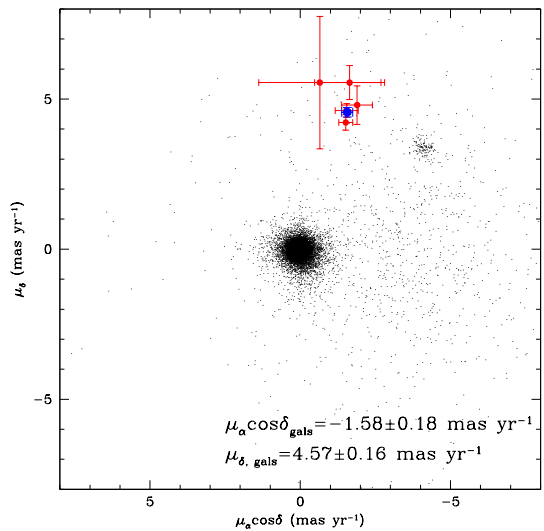}
\caption{\small Position of the five selected galaxies in the VPD of
  the relative PMs. They are shown as red dots. The errorbars
  correspond to the uncertainty of their motions. Their weighted mean
  position is shown as a blue dot together with its uncertainties and
  represents the adopted zero point of the absolute-motion reference
  frame.}
\label{gal_pm}
\end{figure}

\begin{figure}[!htb]
\includegraphics[width=\columnwidth]{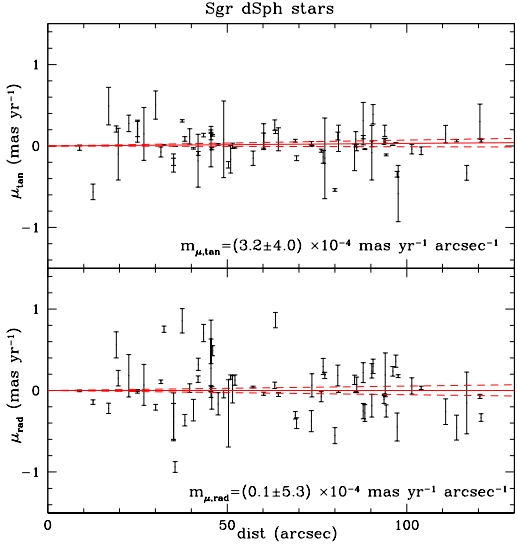}
\caption{\small {\it Upper panel}:\ distribution of the tangential PM
  component of the Sgr dSph members vs. their angular distance from
  the NGC~6681 center.  The best fit of the functional form $f=mr$ is
  shown as a red solid line. The $1\sigma$ uncertainties are traced as
  dashed red lines. The labels indicate the value of the best fit
  slope.  {\it Lower panel}:\ the same for the radial PM component.}
\label{introt}
\end{figure}

\section{RELATIVE PROPER MOTIONS}

The first step in measuring relative PMs was to astrometrically relate
each exposure to a distortion-free reference frame, which from now on
we will refer to as the master frame. We chose the public catalog of
NGC~6681 provided by the ACS survey of Galactic GCs
(\citealt{sarajedini}, \citealt{jay08}) as the master frame.  
We re-scaled this to the UVIS pixel-scale (40 ${\rm mas\, pixel^{-1}}$) for convenience.
We transformed the measured position of each star in each exposure into 
the reference frame by means of a six-parameter linear transformation 
based on the positions of member stars in the reference frame and the 
individual frames.  To maximize the accuracy of these transformations we
treated each chip of our exposures separately, in order to avoid spurious
effects related to the presence of the gap.

As a second step, we selected a sample of reference stars with respect
to which our PMs would be computed.  For convenience, we
chose to compute all PMs relative to the mean motion of the
cluster. Therefore our reference list is based on likely cluster
members. These were initially selected on the basis of their
location on the CMD. We included in the list only well-measured,
unsaturated stars. Then, for each star in each catalog, we computed
the position on the master frame using a transformation based on only 
the closest 50 reference stars.

At the end of the process, for each star we have up to $8$ first-epoch
positions and up to $22$ second-epoch positions on the master frame. To
estimate the relative motion of each star we adopted a
3$\sigma$-clipping algorithm and computed the median X and Y positions of
each star in the first and in the second epoch. The difference between the two 
median positions gives the star's X and Y displacements in $\Delta {\rm T}=5.464$ years.  
The errors in each direction and within each epoch ($\sigma_{1,2}^{{\rm X,Y}}$) 
were computed as the rms of the positional residuals about the median value, 
divided by the square root of the number of measurements $N_{1,2}$:\ ${\rm
  rms_{1,2}}/\sqrt{N_{1,2}}$. Therefore, the error in each PM-component 
associated to each star is simply the sum in quadrature between first- and
second-epoch errors:\ $\sigma_{\rm
  PM}^{{\rm X}}=\sqrt{(\sigma_1^{{\rm X}})^{2}+(\sigma_2^{{\rm X}})^{2}}/\Delta {\rm T}$
and $\sigma_{\rm
  PM}^{{\rm Y}}=\sqrt{(\sigma_1^{{\rm Y}})^{2}+(\sigma_2^{{\rm Y}})^{2}}/\Delta {\rm T}$.

With this first PM determination, we went back to our original
reference-star list and removed those sources whose motion was not
consistent with the cluster's mean motion, (0,0) \masyr by construction.  
We repeated the entire procedure three times, after which the number of
stars in the reference list stopped changing.  To be conservative, we
decided to build the final PM catalog taking into account only the
30$\,$546 stars having at least 3 position measurements in each
epoch. The typical error for well-exposed stars is smaller than 0.002
pixel yr$^{-1}$ in each coordinate, i.e. smaller than 0.07 \masyr\, as
shown in Fig.~\ref{err}.

We converted the PMs into units of \masyr by multiplying the measured
displacements by the pixel scale of the master frame (previously
re-scaled to $0.04 \arcsec/$pixel) and dividing by the temporal
baseline ($5.464$ yrs). Since the master frame is already oriented
according to the equatorial coordinate system, the X PM-component
corresponds to that projected along (negative) Right Ascension
($-\mu_{\alpha}\cos\delta$), while the Y PM-component to that along
Declination ($\mu_{\delta}$).  The output of this analysis is
summarized in Figure \ref{relpm}, where in the upper panels we show
the Vector Point Diagrams (VPDs) and in the lower panels the
corresponding CMDs. Close inspections of the VPDs suggest that at
least three populations with distinct kinematics can be identified in
the direction of NGC~6681.
\begin{enumerate}
\item The cluster population is identified by the clump of stars at
  (0,0) \masyr. By selecting stars within the blue circle in the
  second upper panel, a clean CMD of the cluster is obtained (second
  lower panel of Fig.~\ref{relpm}).
\item A secondary clump of stars is located at roughly $(-4,3.5)$
  \masyr.  Stars selected within the red circle in the third upper
  panel of Fig.~\ref{relpm} define in the CMD (the third lower panel)
  a sequence significantly fainter than that defined by cluster stars
  (as already noticed in the previous Section).  Therefore, these
  stars belong to a population that is both kinematically and
  photometrically different from that of the cluster.  This population
  appears uniformly distributed across the FoV of our observations and
  thus it can be associated to the Sgr dSph, whose main body is
  located in the background of NGC~6681.
\item A much sparser population of stars is centered around $(-4.5,0)$
  \masyr.  The bulk of this population is highlighted with the green
  circle in the last upper panel of Fig.~\ref{relpm}.  The
  corresponding CMD suggests that this is essentially due to
  fore/background sources.
\end{enumerate}

\section{ABSOLUTE PROPER MOTION DETERMINATION}

In this section we describe how we determined the absolute reference 
frame zero point in order to bring our relative PMs (Section 3) into an
absolute system.

\subsection{Absolute reference frame}

In order to measure absolute PMs, an absolute zero point is required.
The best option to define this zero point is to use extragalactic
sources, since they are essentially stationary on account of their enormous
distances.
This method has already been adopted in several previous
works, such as \cite{dinescu99}, \cite{bellini10}, or \cite{sohn12}.
In order to find extragalactic sources we first tried to use the Nasa
Extragalactic Database but found that it is incomplete in the innermost
regions of dense stellar systems like GCs, and provides no detectable
sources in our FoV.  We then performed a careful visual inspection of
our images. Thirty-one galaxies were identified by eye, but only 11 of
them have point-like nuclei and thus are successfully fitted by the
adopted PSF.  Out of these, we selected only the 5 galaxies with an
associated QFIT value (see \citealt{jayking06} for details) smaller
than $0.6$:\ this was necessary to guarantee a measurement of the
source centroid accurate enough to provide a precise determination of
the zero point for the absolute PMs.  Figure \ref{gal} shows how these
galaxies appear in the F814W band.

The selected galaxies are located very close to each other in the
relative-PM VPD (Fig.~\ref{gal_pm}), as expected for distant
sources.  Therefore, we defined the zero-point of the absolute
reference frame as the weighted mean of their relative PMs (see the
blue dot in Figure \ref{gal_pm}):
\begin{equation}
  (\mu_{\alpha}\cos\delta, \mu_{\delta})_{{\rm gals}}=(-1.58\pm0.18, 4.57\pm0.16)~{\rm mas\, yr^{-1}},
\label{eqabsolute}
\end{equation}
as measured with respect to the mean NGC~6681 motion derived in
Section 5.1. The uncertainties correspond to the error on the
calculated weighted means. In order to check whether the quoted 
uncertainties could be underestimated by taking into account only the 
individual PM errors, we computed the reduced $\chi^{2}_{\nu} \equiv \chi^2/(N-1)$ 
of the scatter of the five galaxies around their weighted mean. 
The resulting values are $\chi^{2}_{\nu}=0.17$ and 
$\chi^{2}_{\nu}=1.29$ for the two PM components respectively. 
These values suggest a reasonable estimate for the $\mu_{\delta}$ component 
uncertainty, and a possible overestimate of the $\mu_{\alpha}\cos\delta$ 
component error. However, given the small sample of galaxies, the $\chi^{2}$ 
statistics could be not fully reliable. Therefore we maintained the quoted 
uncertainties throughout the following analysis.

\subsection{Systematic error estimates}

Since any measurement of absolute PMs relies on the accuracy of the
absolute reference frame determination, it is important to look for
possible sources of systematic errors and, if any is found, to quantify their
impact.  One of these sources could be the possible rotation of
NGC~6681 on the plane of the sky.  
Because we used only cluster stars to define the (linear) transformation
between each exposure and the reference frame, if the cluster is rotating,
then our frame will also be rotating.  As such, if NGC~6681 has any component
of rotation in the plane of the sky, our procedure would have introduced 
an artificial rotation
(equal in modulus but opposite in sign) to background and foreground
objects around the cluster center.

NGC~6681 shows a very small ellipticity ($\epsilon=0.01$;
\citealt{harris}, 2010 edition).  Hence, from the relationship between
ellipticity and the rotational parameter $v_{{\rm rot}}/\sigma$
($v_{{\rm rot}}$ and $\sigma$ being, respectively, the cluster
rotational velocity and velocity dispersion; \citealt{illi77}), 
we can reasonably expect that $v_{{\rm rot}}/\sigma \ll0.5$ 
at $r=r_{{\rm h}}$, with $r_{{\rm h}}$ being the half-light radius, 
at least for the edge-on component of rotation.  If the face-on and
edge-on components are similar, we can calculate an upper limit to 
the rotational velocity of
$v_{{\rm rot}}\ll1.4 \times 10^{-3} {\rm mas\, yr^{-1}}{\rm arcsec^{-1}}$,
using $\sigma=5.2$ km$\,$s$^{-1}$, $r_{{\rm h}}=0.71 \arcmin$
and the distance $d=9$ kpc (all from \citealt{harris}, 2010 edition).
This value corresponds to a rotational PM of 0.084 \masyr at a distance 
of $1\arcmin$ from the cluster center.  This is comparable to the random
errors in our absolute astrometry (eq.~[\ref{eqabsolute}]). Hence we
need to carefully check for the possible presence of a rotational
component on the plane of the sky.  To this end we selected Sgr dSph
members since their small PM dispersion (compared to field stars)
produces more stringent limits on the measured rotation values. Note
that the Sgr dSph does not rotate (\citealt{pena}), at least it should
not rotate around the center of NGC~6681. Thus any possible rotation 
signal would
belong to the cluster.

We selected Sgr dSph stars from the CMD (in the intervals $17.5<${\it
  m}$_{{\rm F555W}}<23$ mag and $1.1<${\it m}$_{{\rm F555W}}-${\it
  m}$_{{\rm F814W}}<1.8$ mag) and from the VPD, rejecting sources that
lay beyond $1$ \masyr from its PM bulk distribution.  In addition, we
rejected stars with a PM uncertainty larger than $0.5$ \masyr. Seventy
three stars survived these selection criteria.  We decomposed their PM
vectors into radial ($\mu_{{\rm rad}}$) and tangential ($\mu_{{\rm
    tan}}$) components with respect to the center of NGC~6681.  The
resulting distributions of the $\mu_{{\rm rad}}$ and $\mu_{{\rm tan}}$
components as a function of the distance $r$ from the cluster's center
are shown in Figure \ref{introt}.  We fitted these distributions with
a straight line, forced through the origin (functional form $f=mr$)
since at $r=0$ any internal cluster mean motion (rotation in the PM
tangential component and contraction/expansion in the radial
direction) is zero. We determined the angular coefficient $m$ by 
calculating the minimum of the function d($\chi^{2}$)/d$m$ and we defined 
the errors using the $m$-values corresponding to $\chi^{2}=\chi_{min}^{2}+1$. 
In this approach, we defined $\chi^2 = \sum (\mu_{\rm obs} - mr)^2 / \sigma^2$, 
where $\sigma$ is the scatter of the points around the best fit 
($\sim 0.3$ mas yr$^{-1}$). Hence, the inferred errors on $m$ take into 
account the observed scatter (which includes contributions from the internal 
velocity dispersion of the Sgr dSph stars), and not merely the formal random 
errors on the individual PM measurements.

The best fits of the two distributions give the following slope
values:\ $m_{\mu,{\rm rad}}=(0.1\pm5.3)\times10^{-4}$ \masyrarc and
$m_{\mu,{\rm tan}}=(3.2\pm4.0)\times10^{-4}$ \masyrarc. The fits are
shown in Figure \ref{introt} with red solid lines (the $\pm\sigma$
uncertainty in the fit is shown as red dashed lines).  Our estimated rotation
signal ($\mu_{{\rm tan}}=r\times(3.2\pm4.0)\times10^{-4}$ \masyrarc)
corresponds to $v_{{\rm rot}}=0.82\pm1.02$ km$\,$s$^{-1}$ at $1
\arcmin$ from the cluster center, fully consistent with zero. This
firmly demonstrates that the cluster rotation is negligible in the
plane of the sky. This contrasts with the case of e.g. $\omega$~Cen
(\citealt{vandeven06}).

\cite{ak03} (AK03) estimated the rotation of 47~Tuc from the difference 
of the motion of the Small Magellanic Cloud (located behind the cluster) in two 
pointings on opposite sides of the cluster.  Similar to AK03, in this work we 
have used stars in a background association as a zero-rotation reference to 
obtain very precise limits on the plane-of-sky rotation of a globular cluster.  
This is the first time, however, that such a measurement has been made using 
only one observed pointing at the cluster center.  We want to stress that this 
has only been possible thanks to the small dispersion of the Sgr dSph PMs, which 
translates into a high accuracy in the determination of a possible rotation 
signal.

Since the rotation turned out to be consistent with zero to within 
the uncertainties, we
can assume that PMs of background galaxies are not affected by any
cluster-rotation effect.  To be conservative, however, we used the
uncertainty on the slope $m_{\mu,{\rm rad}}$ to quantify the possible
systematic error on the determination of the absolute reference
frame. We computed the radial and tangential components of the PMs of
the 5 selected galaxies and added to the tangential motion of each
galaxy a term $b=\pm4\times10^{-4}\times r$ (where ${\rm r}$ is
the distance of each galaxy from the cluster center).  Then we
re-calculated the weighted-mean relative PM value, which defines the
absolute zero point.  The difference with respect to the previous
determination is $\Delta|\mu_{\alpha}\cos\delta|=0.023$ \masyr and
$\Delta|\mu_{\delta}|=0.026$ \masyr. These systematic errors on the
determination of the absolute reference frame are significantly
smaller than the random errors (eq.~[\ref{eqabsolute}]), and we
therefore ignore them in the following.

It is also important to be aware of possible contributions from 
parallax to measured positional shifts between observations taken at 
different epochs.  Compared to distant background sources, a foreground 
object will move annually on a parallactic ellipse of semi-major axis 
length $p \equiv 0.1$ (10 kpc / d) mas. 
The positional shift thus introduced between two observations at random times 
separated by a baseline $\Delta T$ is at most twice this value. 
This yields an apparent PM of size $|\Delta PM | \leq 2 p / \Delta T$. 
For our NGC~6681 observations, ${rm d}=9$ kpc and $\Delta T = 5.464$ yr, so that 
$|\Delta PM | \leq 0.04$ mas yr$^{-1}$. This is well below the uncertainties 
in  our absolute reference frame. So while it is not difficult to correct for 
parallax explicitly, we have not done so in the present context. The Sgr dSph 
is farther away than NGC~6681, so any systematic errors in its PM due to 
parallax would be correspondingly smaller.

\begin{figure*}[!htb]
\includegraphics[width=16cm]{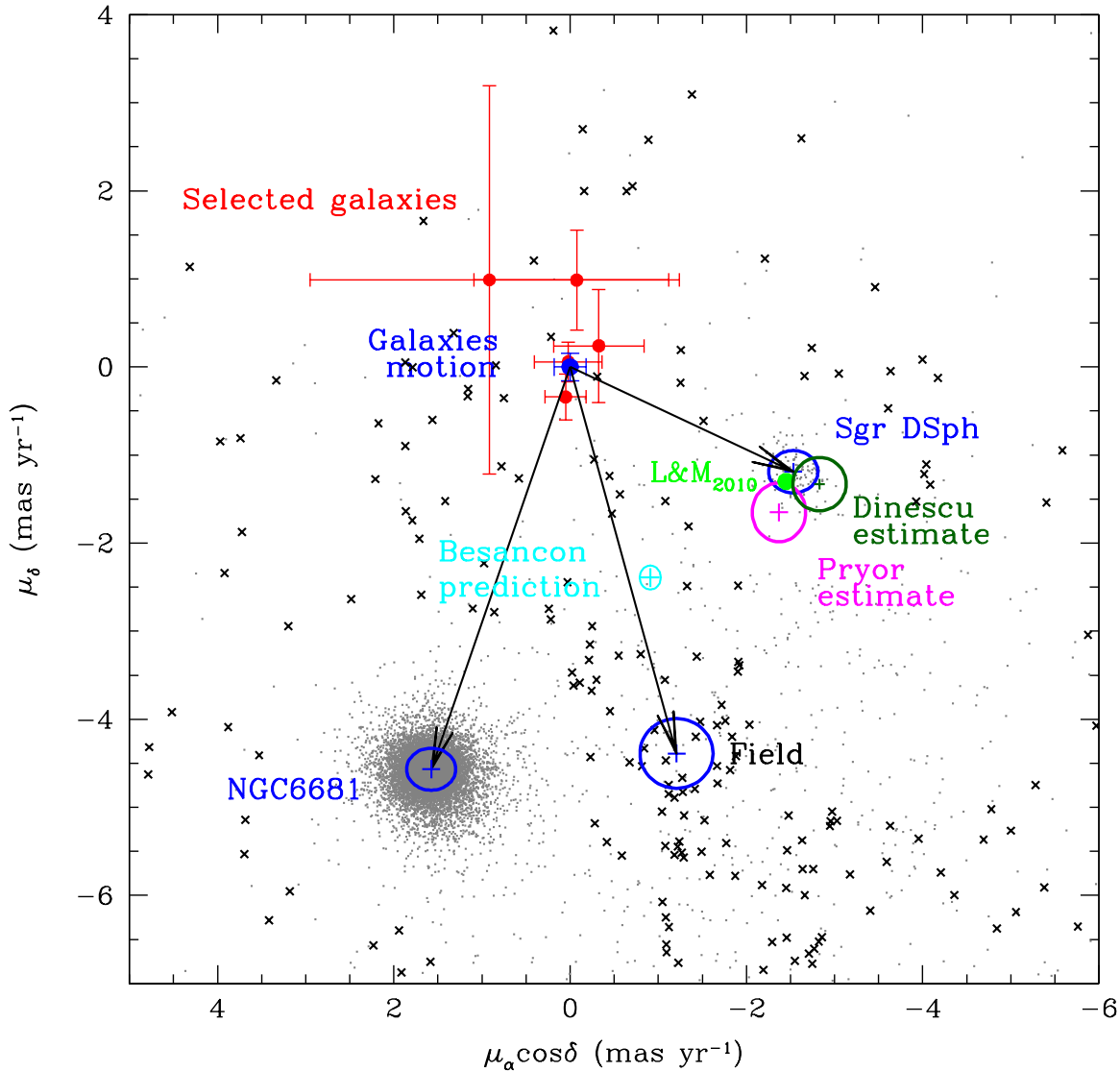}
\caption{\small VPD of the absolute PMs. The red dots indicate the
  selected background galaxies (see also Fig.~\ref{gal_pm}), whose
  mean motion corresponds to the zero point of the VPD. The blue
  ellipses are centered on the measured absolute PMs of the three
  populations (marked with a blue cross) and their size corresponds to
  the calculated 68.3\% confidence region. The black arrows indicate
  their absolute PM vectors.  In the proximity of the Sgr dSph
  estimate, the PM value predicted by \cite{lm10a} is shown as a light
  green dot, while the \cite{pryor10} and \cite{dinescu05}
  measurements and their 68.3\% confidence regions are shown as
  magenta and dark green ellipses, respectively. Finally, the cyan
  ellipse describes the prediction on the PM of the field population
  by the Besan\c{c}on model, which differs from our estimate obtained
  using the stars in the same magnitude and color range (marked with
  black crosses, see the text for the selection criteria).}
\label{absolute}
\end{figure*}

There are many other possible sources of systematic error that can
affect absolute astrometry with {\it HST}. {\it HST} is a very stable instrument,
and most authors use analysis software that is based on the concepts
in \cite{jayking06}. Therefore, most systematic errors should be
similar in magnitude between different studies.  \cite{sohn12,sohn13}
achieved systematic errors $\lesssim 0.03$ \masyr in studies of M31
and Leo~I using techniques that are very similar to those used
here. This is significantly smaller than the random errors in our
absolute reference frame (see eq.~[\ref{eqabsolute}]).  The fact that
our results are not significantly affected by unknown systematics is
also supported by comparison of our PM results with those of other
authors (see Section~5.2 below), which show good agreement to within
the random uncertainties.

\section{ABSOLUTE PROPER MOTION RESULTS}

In this section we present the results for the absolute PMs of the
three populations under investigation.  Each measurement quoted in the
following subsections refers to Figure \ref{absolute}, where the blue
crosses correspond to the absolute PM estimate of each population and
the blue ellipses to their uncertainty.

\subsection{NGC~6681}

In order to measure the absolute PM of NGC~6681 we selected only stars
$1.0$ \masyr from the cluster mean motion and in the magnitude
interval $17.5<${\it m}$_{{\rm F555W}}<22.5$ mag. We iteratively
refined the selection by applying a $3\sigma$ rejection and
re-calculating the barycenter of the PMs as the weighted mean value of
the PMs of the selected stars, until the difference between two
subsequent steps was smaller than $0.01$ \masyr. After the last iterative 
step, a total of $N_{clu}=14\,030$ stars survived the selection
criteria.  We used the sum in quadrature between each single measurement
error and the velocity dispersion of the cluster $\sigma_{v}=0.12$ mas/yr
(based on the line-of-sight velocity dispersion and distance given by \citealt{harris}) 
as weights. To estimate the error $\Delta PM$ on the weighted mean PM in 
each coordinate we use the standard error-in-the-mean, i.e., the dispersion of 
the surviving stars around the weighted mean PM, divided by $\sqrt(N_{\rm clu}-1)$. 
This includes scatter from the internal dispersion of NGC~6681 stars, which 
therefore does not need to be estimated explicitly. We find that the resulting 
error $\Delta PM$ is negligible compared to the error on the absolute reference 
frame. Therefore, the latter dominates the uncertainty on the final absolute PM 
of NGC~6681, which is:\
\begin{equation}
 (\mu_{\alpha}\cos\delta, \mu_{\delta})=(1.58\pm0.18, -4.57\pm0.16)~
  {\rm mas\, yr^{-1}}.
\end{equation}

The PM derived here can be combined with the known distance and
line-of-sight velocity of NGC~6681 from \cite{harris}, to determine
the motion of the cluster in the Galactocentric rest frame. Using the
same formalism, conventions, and solar motion as in \cite{marel12},
this yields $(V_X,V_Y,V_Z) = (203 \pm 2, 111 \pm 9, -179 \pm 7)$
km$\,$s$^{-1}$. This corresponds to a total Galactocentric velocity $|{\vec V}|
= 292 \pm 5$ km$\,$s$^{-1}$. This significantly exceeds the central velocity
dispersion $\sigma \approx 120$ km$\,$s$^{-1}$ of the Milky Way's spheroidal
components (e.g., \citealt{deason12}). Hence, NGC~6681 must spend most
of the time along its orbit at significantly larger distances from the
Galactic Center than its current distance of $2.2$ kpc
(\citealt{harris}).

\subsection{Sagittarius Dwarf Galaxy}

In order to determine the absolute PM of the Sgr dSph we basically
followed the same procedure previously described for NGC~6681.  
In setting the weights for the PM averaging, we used the dispersion 
$\sigma_v \sim 0.3$ \masyr implied by Figure 6. This includes both 
contributions from the internal velocity dispersion of the Sgr dSph (see
e.g. \citealt{frinchaboy12}), and unquantified systematic errors.
We selected stars within $1.0$ \masyr from the Sgr dSph mean motion and
in the interval $17.5<${\it m}$_{{\rm F555W}}<23.5$, which is one
magnitude fainter with respect to the case of NGC~6681, since most of
the Sgr dSph stars belong to its faint MS. The resulting absolute PM
is:\
\begin{equation}
 (\mu_{\alpha}\cos\delta, \mu_{\delta})=(-2.54\pm0.18,
  -1.19\pm0.16)~{\rm mas\, yr^{-1}}.
\end{equation}

We compared this value with previous estimates.  With the aim of
reconstructing the kinematical history of this galaxy and to predict
its evolution in a triaxial Milky Way halo, \cite{lm10a} built a
N-body model able to reproduce most of the system's observed
properties. In the Law \& Majewski model, the Sgr dSph
has a Galactocentric motion $(V_X,V_Y,V_Z) = (230, -35, 195)$ km$\,$s$^{-1}$,
corresponding to a total velocity $|{\vec V}| = 304$ km$\,$s$^{-1}$.
The absolute PM predicted by the model is
($\mu_{\alpha}\cos\delta, \mu_{\delta})=(-2.45, -1.30$) \masyr (light
green dot in Figure \ref{absolute}).  An estimate of the
absolute PM of the Sgr dSph based on {\it HST} data has been recently presented by
\cite{pryor10}. The authors used foreground Galactic
stellar populations as reference frame and they determined an absolute
PM of ($\mu_{\alpha}\cos\delta, \mu_{\delta})=(-2.37\pm0.2,
-1.65\pm0.22$) \masyr, which is shown as a magenta ellipse in Figure
\ref{absolute}.  A ground-based estimate of the absolute PM of the Sgr
dSph was presented by \cite{dinescu05}. Using the Southern Proper
Motion Catalog 3 they determined that $(\mu_\alpha \cos\delta,
\mu_\delta) = -2.83 \pm 0.20, -1.33 \pm 0.20)$ \masyr, which is shown
as the dark green ellipse in Figure \ref{absolute}.  These previous
estimates are in rough agreement with the value determined here.

It is worth noting, however, that these other determinations are not
directly comparable with ours, since they refer to different regions
of the Sgr dSph. Indeed, this has two possible effects. The first one
is that possible internal motions, such as rotation, could translate
into different mean motions, thus introducing a systematic effect.
This should not be a problem for the Sgr dSph, since this galaxy does
not show any evidence of rotation (\citealt{pena}).  The second effect
is that if the whole galaxy has a 3D velocity vector different from
zero, then the observed PMs for different pointings are not the same,
because of perspective effects due to the imperfect parallelism
between the lines of sight (\citealt{marel02}).  Since the Sgr dSph is
a nearby galaxy, this effect could be relevant and we calculated the
correction to apply (as in \citealt{marel08}) in order to obtain
comparable estimates at the center of mass of the Sgr dSph.

Under the hypothesis that the center of mass of the Sgr dSph is moving
as the \cite{lm10a} prediction, our perspective-corrected PM
measurement becomes ($\mu_{\alpha}\cos\delta,
\mu_{\delta})=(-2.56\pm0.18, -1.29\pm0.16$) \masyr. The corrected
\cite{pryor10} estimate becomes ($\mu_{\alpha}\cos\delta,
\mu_{\delta})=(-2.37\pm0.20, -1.63\pm0.22$) \masyr, and the corrected
\cite{dinescu05} estimate becomes $(\mu_\alpha \cos\delta, \mu_\delta)
= -2.83 \pm 0.20, -1.56 \pm 0.20)$ \masyr. Thus our measurement is
consistent with the previous observations. The weighted average of all
observational estimates of the center-of-mass PM of the Sgr dSph is
$(\mu_\alpha \cos\delta, \mu_\delta) = (-2.59 \pm 0.11, -1.45 \pm
0.11)$ \masyr. This is consistent with the theoretical model of
\cite{lm10a}, once the uncertainties on transforming that into a PM
value (e.g., from uncertainties in the distance and solar motion) are
taken into account as well.  Therefore, our
measurement is consistent within about a $1\sigma$ uncertainty both with
theoretical predictions (\citealt{lm10a}) and the previous {\it HST}
observations (\citealt{pryor10}).

\subsection{Field}

We compared the absolute PMs of Field stars in our catalog with those
predicted in the same region of sky by the Besan\c{c}on Galactic model
(\citealt{robin}). We generated a simulation over a 0.01 square
degrees ($6\arcmin\times6\arcmin$) FoV around the center of NGC~6681
($l=2$\textdegree\!.$85$, $b=-12$\textdegree\!.$51$) and 50 kpc deep.
To minimize any possible bias, we have constructed a sample as similar
as possible to the observed stars, based on a comparison between the
observed and the simulated CMDs.  Simulated field stars were selected
within the magnitude range:\ $17.5<${\it m}$_{{\rm F555W}}<22.5$ mag
and ({\it m}$_{{\rm F555W}}-${\it m}$_{{\rm F814W}}$)$>1.5$ mag and
1378 stars survived these criteria. 
The average predicted motion is shown in Figure \ref{absolute} as a
cyan ellipse, which corresponds to ($\mu_{\alpha}\cos\delta,
\mu_{\delta})=(-0.91\pm0.08, -2.39\pm0.09$) \masyr.\footnote{It 
would be easy to reduce the random uncertainty on this model prediction 
by drawing a larger number of simulated stars. However, we have not 
pursued this since the accuracy of the prediction is dominated largely 
by systematic errors in the model assumptions anyway.}

Field stars in our observed catalog were selected following the same
color and magnitude cuts. We also required these stars to have PM
errors smaller than $0.2$ \masyr in each coordinate. Finally, we
excluded those stars within $1.8$ \masyr of the cluster mean motion
and within $1.0$ \masyr of the Sgr dSph mean motion.  We iteratively
removed field stars in symmetric locations with respect to the Sgr dSph
and NGC~6681 exclusions in order to better define the mean motion of
the Field population and adjusted the weighted mean motion after each
iteration (thus following the method described by \citealt{jay10} for
the determination of the center of $\omega$~Cen).  The 281 selected
field stars used for the final estimate are shown as black crosses in
Figure \ref{absolute}.  Since these stars display a large 
scatter in the VPD due to their velocity dispersion and not to their random errors, 
in this case we computed a statistically more appropriate $3\sigma$-clipped 
unweighted mean motion. It is
shown as a blue ellipse in Figure \ref{absolute} and its value is:\
\begin{equation}
 (\mu_{\alpha}\cos\delta, \mu_{\delta})=(-1.21\pm0.27, -4.39\pm0.26)
  {\rm mas\, yr^{-1}}.
\end{equation}

Our PM measurement is similar to the prediction of the Besan\c{c}on
model, in that it points in the same direction on the sky (see
Figure~\ref{absolute}). However, the sizes of the PM vectors are not
formally consistent to within the random errors.  Since our
measurements for Sgr dSph stars are entirely consistent with both
previous measurements and theoretical predictions, this cannot be due
to systematic errors in our measurements (which would affect all point
sources equally). Instead, the mismatch is most likely due to
shortcomings in the Besan\c{c}on models. In particular, for pointings
this close to the Galactic Plane, the model predicted PM distribution
is likely to depend sensitively on the adopted dust extinction model,
which is poorly constrained observationally. Also, the model predicted
PM distribution depends on the solar motion in the Milky Way, which
continues to be debated (e.g., \citealt{mcmillan11};
\citealt{bovy12}).

\section{CONCLUSIONS}

We have analyzed two sets of {\it HST} observations separated by a temporal
baseline of $5.464$ years in the direction of the Galactic GC NGC~6681
in order to obtain the first-ever measurement of its absolute PM, as
well as the absolute motions of the Sgr dSph and the field population
in this direction.  First, we obtained relative PMs for a total of
about 30$\,$000 sources. For the brighter ones, the uncertainties are
smaller than $\sim 0.07$ \masyr in each coordinate.  Then, by using
background galaxies, we determined the zero-point of the
absolute-motion reference frame with an uncertainty of $0.18$ and
$0.16$ \masyr on the $\mu_{\alpha}\cos\delta$ and $\mu_{\delta}$
components, respectively. We also quantified the systematic errors on
the definition of the absolute reference frame due to possible
internal rotation of NGC~6681.  We demonstrated that very stringent
constraints on the rotation of a GC can be obtained by using
non-member populations, provided that their PMs have a sufficiently
small dispersion.  We used Sgr dSph stars as the non-member population and
we estimated a rotational velocity for NGC~6681 of $v_{{\rm
    rot}}=0.82\pm1.02$ km$\,$s$^{-1}$ at $1 \arcmin$ from the cluster
center, consistent with zero. This corresponds to negligible
systematic errors of $(0.023, 0.026)$ \masyr  in the previously quoted
PM components.

We measured the absolute PM for the three populations under
investigation.  The absolute PM of NGC~6681 is
$(\mu_{\alpha}\cos\delta, \mu_{\delta})=(1.58\pm0.18, -4.57\pm0.16)$
\masyr.  We also measured the absolute PM of the Sgr dSph and compared
it with previous determinations and model predictions. Our estimate is
$(\mu_{\alpha}\cos\delta, \mu_{\delta})=(-2.57\pm0.18, -1.14\pm0.16)$
\masyr.  After correction for viewing perspective to obtain an
estimate of the PM of the Sgr dSph center of mass, this value is
consistent within about a $1\sigma$ uncertainty with both the model
prediction of \cite{lm10a} and the perspective corrected measurements
of \cite{pryor10} and \cite{dinescu05}.  Finally, we estimated the
absolute PM of the field population intercepted along the line of
sight to be $(\mu_{\alpha}\cos\delta, \mu_{\delta})= (-1.17\pm0.27,
-3.99\pm0.26)$ \masyr.

\acknowledgements{ This research is part of the project COSMIC-LAB
  (web site: www.cosmic-lab.eu) funded by the European Research
  Council (under contract ERC-2010-AdG-267675).  Support for this work
  was provided by NASA through a grant for pro- gram AR-12845 from the
  Space Telescope Science Institute (STScI), which is operated by the
  Association of Universities for Research in Astronomy (AURA), Inc.,
  under NASA contract NAS5-26555.}

\end{document}